\documentclass[%
 reprint,
 amsmath,amssymb,
 aps,floatfix
]{revtex4-2}

\usepackage{graphicx}
\usepackage{dcolumn}
\usepackage{bm}
\usepackage{hyperref}
\usepackage{xcolor}

\begin{document}

\preprint{APS/123-QED}

\title{{Tidal deformability of dark matter admixed neutron stars}}

\author{Kwing-Lam Leung}
\author{Ming-chung Chu}
\author{Lap-Ming Lin}
\affiliation{Department of Physics and Institute of
Theoretical Physics, \\The Chinese University of Hong Kong,
Shatin, Hong Kong S. A. R., China}

\date{\today}

\begin{abstract}
The tidal properties of a neutron star are measurable in the
gravitational waves emitted from inspiraling binary neutron
stars, and they have been used to constrain the neutron star
equation of state. In the same spirit, we study the
{dimensionless} tidal deformability of dark matter
admixed neutron stars. {The tidal Love number is
computed in a two-fluid framework.} The dimensionless tidal
Love number and dimensionless tidal deformability are computed
for dark matter admixed stars with the dark matter modelled as
ideal Fermi gas or self-interactive bosons. The dimensionless
tidal deformability shows a sharp change from being similar to
that of a pure normal matter star to that of a pure dark
matter star, within a narrow range of intermediate dark matter
mass fraction. Based on this result, we illustrate an approach
to study the dark matter parameters through the tidal
properties of massive compact stars, making use of the
self-similarity of the dimensionless tidal deformability-mass
relations when the dark matter mass fraction is high.
\end{abstract}

\maketitle

\section{\label{sec_intro}Introduction}
Most of the mass in the universe is believed to be dark matter
(DM). However, almost all properties of DM are still unknown,
and the existence of DM is only supported by indirect
evidences \cite{pp_DM_candidates}. Different ways to study DM
are conducted, such as measurements of rotation curves of
spiral galaxies \cite{pp_spiral_galaxy}, cosmic microwave
background \cite{pp_CMB_DM}, and gravitational lensing
\cite{pp_lensing_DM, pp_lensing_DM2}. Searching for DM particles is also
ongoing in different experiments
\cite{pp_SHIP_phys, pp_XENON10, pp_XENON}. The recent
observation of excess events reported by the XENON1T
experiment \cite{pp_XENON} may be the first direct detection
of DM, and if so, it may open up a window for discovering
physics beyond the Standard Model. Understanding the nature of
DM would be a significant advance of physics.

Since so far only observations through gravity reveal the
existence of DM, perhaps gravity is the only interaction
between the DM and Standard Model particles, or normal matter
(NM). Due to the weak coupling strength of gravitational
interaction, it would be difficult to investigate the DM
through its interaction with NM. It may be easier if we study
the DM in the cosmological scale, where the DM contributes a
large part of gravity. Another possibility is to focus on high
density regions, as gravity plays a significant role there.
Compact stars, such as neutron stars, can be a possible
natural laboratory to study DM.

Due to the high matter density at the neutron star core, the
physics in this region is still not well understood. It is
thus important and interesting to study the properties of
neutron stars, which can be used to constrain the unknown
nuclear matter equation of state (EOS).  
For example, the mass-radius relation and tidal deformability
of neutron stars have been studied
extensively (see, eg.,
\cite{pp_NS_EOS,LATTIMER_2007,Chatziioannou_2020} for
reviews). 
Although calculating the nuclear matter EOS from first
principle is still not possible, nuclear physics experiments
and neutron star observations have given constraints on the
EOS. The recent accurate measurement of the neutron skin
thickness of ${}^{208}$Pb has constrained the density
dependence of the symmetry energy near saturation density
\cite{neutron-skin}. 
The observations of neutron stars with masses 
$\approx 2 M_\odot$ \cite{pp_mass_limit, pp_2solar_pulsar}
have already ruled out many soft EOS models. 
The tidal deformability of neutron stars has also been
constrained by the observation of the first gravitational-wave
event GW170817 from a binary neutron star system
\cite{GW170817}, and implications 
on the EOS models have been studied (e.g.,
\cite{pp_GW170817_EOS,Annala:2018,De:2018,Fattoyev:2018,
Most:2018,Tews:2018,Lim:2018}).
A 2.6 \(M_\odot\) compact object recently observed in a
gravitational-wave event GW190814 \cite{pp_26} will also be a
challenge to our understanding of dense nuclear matter if that
object is a neutron star 
(see, e.g., \cite{Most_2020,Essick_2020,Zhang_2020,
pp_26_spinAndMass,Dexheimer_2021}
for various proposals), though the probability that it is a
black hole is high according to recent studies
\cite{Fattoyev_2020,Tews_2021}. 
The more recent mass-radius measurements of pulsars PSR
J0030$+$0451 \cite{Riley_2019,Miller_2019} and PSR
J0740$+$6620 \cite{Riley_2021,Miller_2021}
obtained by the NICER X-ray telescope have also yielded
important information about the EOS. 
With the prospect of seeing more neutron-star observations in
both the electromagnetic
and gravitational-wave channels, we should be able to gain a
much better understanding 
of the unknown nuclear matter EOS in the coming decade.
Furthermore, neutron stars 
may also be used to probe the nature of DM and help to answer
one of the fundamental questions in physics as mentioned
above.

{Compact objects with DM admixture have been studied
previously, such as supernova progenitors \cite{Chan_2021,
Leung_2015} and neutron stars (see, e.g.,
\cite{pp_NS_DMcore,pp_admixedNS,Xiang_2014,Rezaei_2017,
Ellis_2018,pp_admixedNS_0.05,
Gresham_2019,Deliyergiyev_2019,pp_DMhalo,Das_2020,pp_preprint,
Kain_2021,Lee_2021}). }
With its relevance to the gravitational-wave signals from binary neutron stars,
the tidal deformability of neutron stars with small amount of
DM admixtures has also
been studied in \cite{pp_admixedNS_0.05}, and it was suggested
that a 5\% DM mass fraction in a neutron star can already
alter the conclusion about ruling out neutron star EOSs. The
tidal properties of boson stars \cite{pp_tidal_BS} and pure DM
stars \cite{pp_DMstar} have been studied as well. The tidal
properties of compact stars can be a tool to discover new
classes of compact stars. In this work, we assume that the DM
and NM only couple through gravity. The mass-radius relation
and tidal properties of DM-admixed neutron stars are studied
with a two-fluid treatment.

The plan of the paper is as follows. 
In Section \ref{sec_method}, we describe the formulation to calculate the 
hydrostatic equilibrium and the tidal Love number of DM admixed stars. We also
discuss the EOS models employed for the NM and DM. Our numerical 
results are presented in Section \ref{sec_result} and our conclusions are
summarized in Section \ref{sec_discussion}. Unless otherwise noted, we use
units where $G=c=1$.

\section{\label{sec_method}Method}
\subsection{\label{sec_static}Hydrostatic configuration}
The tidal deformability of a nonrotating compact star is determined by 
perturbative calculations starting from the unperturbed background solution
described by a spherically symmetric and static metric
\begin{equation} 
ds^2 = - e^{\nu(r)} dt^2 + e^{\lambda(r)} dr^2 + r^2 (d\theta^2 + \sin^2\theta 
d\varphi^2 ) .
\end{equation} 
The equilibrium structure of a nonrotating compact star is determined by the 
Tolman-Oppenheimer-Volkoff (TOV) equation \cite{pp_TOV}:
\begin{eqnarray}
\frac{dp}{dr}&=& -\frac{m+4\pi r^3 p}{r^2(1-2m/r)}(\rho+p),\label{eq:dpdr}\\
\frac{dm}{dr} &=& \;4\pi r^2\rho, \\
\frac{d\nu}{dr}&=&   {\frac{ 2 (m+4\pi r^3 p)}{r^2 (1 - 2m/r)} } ,
\end{eqnarray}
where $\rho$ and $P$ are the energy density and pressure, respectively. The
function $m(r)$ is defined by $e^{-\lambda(r)} = 1 - 2m(r)/r$. The TOV 
equation is closed by providing an EOS $p(\rho)$.  
The conditions at the star center are \(m(r=0) = 0\) and 
\(\rho(r=0) = \rho_c\), with \(\rho_c\) a given central density. The TOV
equation will be solved from \(r=0\) to \(R\), where \(R\) is the radius of the
star defined by \(p(R)=0\). The total mass of the star \(M\) will be \(m(R)\).
Taking proper limit of the right-hand side of 
Eq.~(\ref{eq:dpdr}), we have \(dp/dr\rightarrow 0\) when \(r\rightarrow 0\).  
The metric function $\nu(r)$ has the boundary condition $e^{\nu(R)}=1-2M/R$ at
the surface.   

In order to study two-fluid DM-admixed stars, some modifications are needed for
the TOV equation. The energy density in general will depend on both the number
densities of NM and DM. We may express the energy density as 
\begin{align}
\rho(N_n,N_d) = \rho_n(N_n) + \rho_d(N_d) + \rho_{interact}(N_n,N_d),
\label{eq_decouple}
\end{align}
where \(N_i\) is the number density, and \(i=\) \(n\), \(d\) denotes the NM and
DM components, respectively.
The total energy density is the sum of the contributions of each component and
the interaction part \(\rho_{interact}\). In this study, we assume that the NM
and DM only interact through gravity. Therefore, \(\rho_{interact} = 0\), and
\(\rho\) can be separated into two individual parts, each depending only on one
of the components. Thus, the pressure can also be separated into two parts, and
we can define them as the pressure of the NM and DM. {From the analogy
to the Newtonian situation, we can construct a set of equations by considering
the pressure of one component will not support the other component.} We have a
two-fluid version of the TOV equation \cite{pp_NS_DMcore, pp_preprint}:
\begin{eqnarray}
\frac{dp_i}{dr}&=& -\frac{m+4\pi r^3 p}{r^2(1-2m/r)}(\rho_i+p_i),
\label{eq:TOV1} \\
\frac{dm_i}{dr}&=& 4\pi r^2\rho_i, \\
\frac{d\nu}{dr}&=&  { \frac{2 (m+4\pi r^3 p)}{r^2 (1 - 2m/r)} } ,
\end{eqnarray}
where \(i = n\) or \(d\). Variables with a subscript denote the quantities of
the corresponding component, and variables without the subscript denote the sum
of the two components (i.e., $m=m_n+m_d$ and $p=p_n+p_d$).
The conditions at the star center are \(m_i(r=0)=0\) and
\(\rho_i(r=0)=\rho_{c,i}\). The pressure of the two components in general drop
to zero at different \(r\). The radius of the star $R$ is defined to be the
outermost one, where the pressure of both components vanish. The original TOV
equation will be recovered if we add up the two components. {The above
set of hydrostatic equilibrium equations can in fact be derived from a general
relativistic two-fluid formalism \cite{Comer1999} assuming that the two fluids
only interact via gravity (see Appendix \ref{sec_TOV}).}

\subsection{\label{sec_tidal}Tidal Love Number and {Dimensionless} Tidal
Deformability}

The deformation of a compact star due to the tidal effect created by a
companion star
is characterized by the tidal deformability $\lambda_{tid}$ which is defined by
$Q_{ij} = - \lambda_{tid} {\cal E}_{ij}$, where $Q_{ij}$ is the traceless
quadrupole moment tensor of the star and ${\cal E}_{ij}$ is the tidal field
tensor. The computation of $\lambda_{tid}$ for non-rotating neutron stars is
well established. 
Here we only summarize the main equations and refer the reader to
\cite{pp_tidal_NS,Damour_2009,pp_quark} for more details. The linearized metric
and fluid equations yield the following equation for determining a perturbed
metric variable $y(r)$:
\begin{equation} 
ry'+y^2+y e^{\lambda} \left[1+4\pi r^2(p -\rho) \right] +r^2Q=0  ,
\label{eq:Love_y}
\end{equation}
where primes denote radial derivatives and the function $Q(r)$ is given by 
\begin{equation}
Q = 4\pi e^{\lambda }\left( 5\rho +9p +\frac{\rho +p }{dp/d\rho} \right) 
- \frac{6e^{\lambda}}{r^2}-(\nu')^2 .  \label{eq_Q}
\end{equation}
The boundary condition at the center is $y(0)=2$. After matching the interior 
and exterior solutions of Eq.~(\ref{eq:Love_y}) at the surface, one can obtain
the so-called (quadrupolar) tidal Love number $k_2$:
\begin{align}
k_2 = &\;\tfrac{8}{5}\beta^5(1-2\beta)^2[2-y_R+2\beta(y_R-1)]\nonumber\\
&\times\{2\beta(6-3y_R+3\beta(5y_R-8))\nonumber\\
&+4\beta^3[13-11y_R+\beta(3y_R-2)+2\beta^2(1+y_R)]\nonumber\\
&+3(1-2\beta)^2[2-y_R+2\beta(y_R-1)]\log(1-2\beta)\}^{-1},
\label{eq_k2}
\end{align}
where \(\beta = M/R\) is the compactness parameter and \(y_R = y(r=R)\). The
tidal deformability is then given by 
\begin{equation} 
\lambda_{tid} = {\frac{2}{3}} k_2 R^5 . 
\end{equation} 
It is also convenient to define the dimensionless tidal deformability 
$\Lambda = \lambda / M^5$. {In this paper, we only focus on the
dimensionless tidal deformability $\Lambda$, but not $\lambda_{tid}$.} The
weighted average of \(\Lambda\) of a binary neutron system can be inferred from
the gravitation waves emitted during the inspiral phase of the system
\cite{pp_tidal_GW, pp_tidal_BS}. Also, \(\Lambda\) is studied in the I-Love-Q
relation \cite{pp_iloveq}, 
an EOS-insensitive universal relation found for neutron stars.

{For the two-fluid case, some modifications of Eq.s \ref{eq:Love_y} and
\ref{eq_Q} are needed. The energy density, pressure and mass can be replaced by
the two components' sums. The term with $dp/d\rho$ requires some calculations.
We follow the general relativistic two-fluid formalism in \cite{pp_tidal_super}
and derive the modification needed in Appendix \ref{sec_TOV}:}
\begin{align}
\frac{\rho + p}{dp/d\rho} \to \sum_i\frac{\rho_i + p_i}{dp_i/d\rho_i}.
\end{align}
{It should be noted that this is valid only if the NM and DM do not
interact microscopically in the sense that the energy density function can be
decomposed as in Eq.~(\ref{eq_decouple}) with a vanishing interaction part
(i.e., \(\rho_{interact} = 0\), assumed in this paper). For the more general
case, $\rho_{interact} \neq 0$, one can employ the formulation in
\cite{pp_tidal_super}, which was originally developed for two-fluid superfluid
neutron stars (see also \cite{Yeung_2021}). }

\subsection{\label{sec_dmEOS}Equation of State for Dark Matter}
There are many candidates for DM particles, such as axions, sterile neutrinos
and different possible WIMPs \cite{pp_DM_candidates}. 
{Since the nature of DM is uncertain at this point, we consider both
fermionic and bosonic DM particles and use only simple models to represent the
DM EOS.} The two types of EOS we use are zero-temperature ideal Fermi gas and
self-interactive bosons with a quartic term of the scalar field in the
Lagrangian density. Both models can be approximated by polytopic EOSs in some
limits. {The free parameters will be the particle mass, or a combination
of the particle mass and strength of self-interaction.}

\subsubsection{\label{sec_fermiDM}Fermionic Dark Matter}
The first DM model we will use is the zero-temperature ideal Fermi gas. Stars
supported by electron degeneracy pressure is a successful model for white
dwarfs. For a better treatment, the EOS for white dwarfs may also include the
contribution from the Coulomb force. The first modeling of neutron stars was
done similarly by using a zero-temperature ideal neutron gas EOS \cite{pp_TOV}.
Although we now know that the neutron star EOS is much more complicated, this
attempt still gives the right orders of magnitude for different properties of
neutron stars.

We assume there is only one type of spin-1/2 DM particles. The zero-temperature
ideal Fermi gas EOS \cite{pp_TOV} is
\begin{align}
\rho =& \:K(\text{sinh}\:t - t),\\
p =& \:\frac{1}{3}K(\text{sinh}\:t-8\:\text{sinh}\:\frac{1}{2}t + 3t),
\end{align}
with
\begin{align}
K &= \frac{\pi\mu^4}{32\pi^3\hbar^3},\\
t &= 4\ln[y+(1+y^2)^{1/2}],
\end{align}
where
\begin{align}
y = \left(\frac{3\pi^2\hbar^3n}{\mu^3}\right)^{1/3},
\end{align}
\(\mu\) is the particle mass, and \(n\) is the number density. In the
non-relativistic and ultra-relativistic limits, the EOSs become polytopic with
indices \(\frac{3}{2}\) and \(3\), respectively.

\subsubsection{\label{sec_bosonDM}Bosonic Dark Matter}
Unlike the fermionic case, bosons do not have degeneracy pressure. To have a
bosonic DM component which can be supported against the gravity,
self-interaction for the bosonic DM is assumed, which can be modeled in a
simple way. We follow the method in \cite{pp_bosonEOS, pp_bosonEOS_limit},
which add an additional quartic term of the scalar field in the Lagrangian.
When the ratio \(am_{Planck}^2/4\pi\mu^2\) is large, an effective EOS for this
self-interacting bosonic DM \cite{pp_bosonEOS} is 
\begin{align}
p = \frac{4}{9}\rho_0[(1+\frac{3}{4}\rho/\rho_0)^{1/2}-1]^2 ,
\label{eq:boson_EOS}
\end{align}
where
\begin{align} 
\rho_0 = \frac{\mu^4}{4a\hbar^3},
\end{align}
$\mu$ is the particle mass, 
\(a\) is a dimensionless constant describing the strength of the
self-interaction, and \(m_{Planck}\) is the Planck mass,
\begin{align} 
m_{Planck} = \sqrt{\frac{\hbar c}{G}}.
\end{align}
In the low and high density limits, the EOS reduces to the following polytropic
forms:
\begin{align}
&p = \frac{\rho^2}{16\rho_0},& \text{ for low density,}
\end{align}
\begin{align}
&p = \frac{1}{3}\rho,& \text{ for high density}.
\end{align}

\subsection{\label{sec_nmEOS}Equation of State for Normal Matter}
From the gravitational-wave signals of the GW170817 event, nuclear matter EOS
is constrained and ``soft" EOSs such as the APR EOS are favored over ``stiff"
ones \cite{pp_GW170817_EOS}. However, the APR EOS cannot account for the 2.6
$M_\odot$ object of the GW190814 event
\cite{pp_26}, if it is a neutron star. So, we use the APR EOS \cite{pp_APR} for
NM and study how the DM admixture may affect the result. 
For comparison, the Skyrme model parameterizations \cite{pp_Skyrme} of LNS EOS
\cite{pp_LNS} and KDE0v1 EOS \cite{pp_KDE0v1} are used for the NM as well.
These two EOSs have a maximum stellar mass (\(M_{max}\)) less than 2
\(M_\odot\), and they would be ruled out by the
$2 M_\odot$ observational constraint \cite{ pp_mass_limit, pp_2solar_pulsar}. 
However, DM-admixed neutron stars constructed with these EOSs may reach a
larger $M_{max}$ than the usual pure NM neutron stars as the DM component is
included.

\subsection{Properties of Pure NM Neutron Stars and Pure DM Stars}

\begin{figure}[htb]
\includegraphics[width=\linewidth]{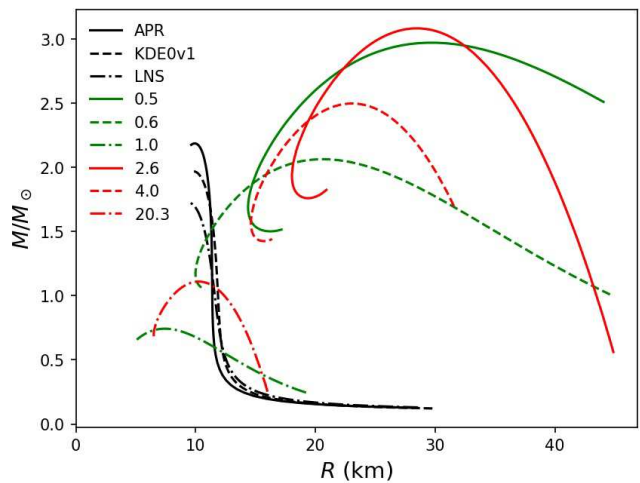}
\caption{\label{fig_m-r_all} Mass-radius relations for different compact stars.
Pure NM neutron stars (black lines) are modeled by the APR, KDE0v1, and LNS
EOSs. Fermionic DM stars (green lines) modeled by ideal Fermi gas EOS are
labeled by the particle mass $\mu$ (in GeV). Bosonic DM stars (red lines) are
labeled by $\rho_0 \hbar^3$ (in $10^{-4}$ GeV\(^4\)). }
\end{figure}

Before studying the properties of DM admixed neutron stars, we consider the
structures of 
pure NM neutron stars and DM stars for our EOS models. For the fermionic DM,
the particle mass is chosen to be in the order of \(O(0.1)\) GeV, so that the
constructed pure DM star will have a mass in the order of solar mass. For the
bosonic DM, \(\rho_0\hbar^3\) is chosen to be in the order of \(O(10^{-4})\)
GeV\(^4\), which also generates a pure boson star in solar mass scale. Note
that our choices of EOSs and parameters for NM and DM are just limiting cases
to illustrate the situation before admixing the two components. 
Readers may refer to \cite{pp_NS_EOS_Mass,pp_DMstar} for more discussion of the
nuclear matter and DM EOSs. 
The pure DM stars generated with these parameter values have radii and masses
comparable to typical neutron stars. 
The mass-radius relations for various EOSs are shown in Fig. \ref{fig_m-r_all}.
We find that the ideal Fermi gas and the self-interactive boson EOSs behave
self-similarly under different choices of parameters, as there are
dimensionless solutions for these EOSs \cite{pp_DMstar}. Results scale with
some combination of the DM parameters. For pure fermionic DM stars, \(M_{max}
\propto \mu^{-2}\), and \(M_{max}\propto\rho_0^{-1/2}\propto\sqrt{a}\mu^{-2}\)
for pure bosonic DM stars \cite{pp_DMstar}. So, the \(M_{max}\) of fermionic DM
stars depends sensitively on the DM particle mass, increasing by around 1
\(M_\odot\) when $\mu$ is decreased from 0.6 GeV to 0.5 GeV.

\begin{figure}[htb]
\includegraphics[width=\linewidth]{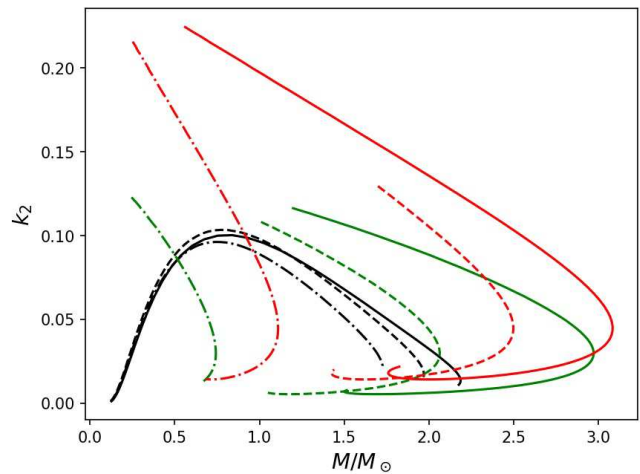}
\caption{\label{fig_k2-m_all} Tidal Love number against total mass for the same
EOSs and parameters as those in Fig. \ref{fig_m-r_all} .}
\end{figure}

\begin{figure}[htb]
\includegraphics[width=\linewidth]{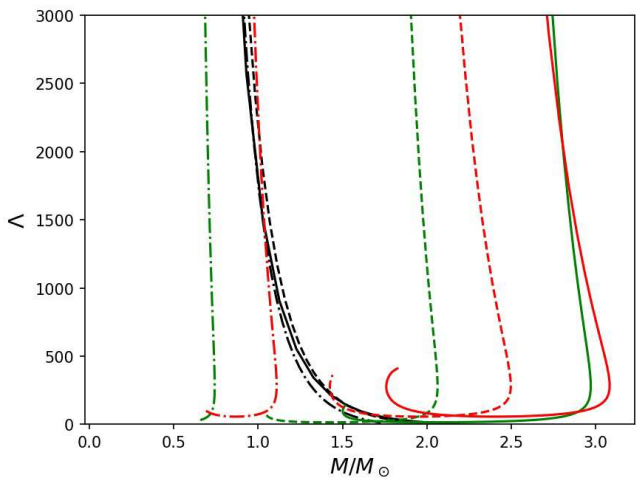}
\caption{\label{fig_lambdabar-m_all}Same as  Fig. \ref{fig_k2-m_all}, but for
the dimensionless tidal deformability.}
\end{figure}

The tidal Love number and dimensionless tidal deformability of the stars shown
in Fig. \ref{fig_m-r_all} are plotted against the total mass \(M\) in Fig.
\ref{fig_k2-m_all} and Fig. \ref{fig_lambdabar-m_all}, respectively. These
plots give us some understanding about each EOS. Indeed, the \(\Lambda-M\)
relation normalized by the \(M_{max}\) of each curve is independent of the
particle mass, for both fermionic and bosonic DM. The dimensionless tidal
deformability is sensitive to \(\mu\) (\(\rho_0\)) for the fermionic (bosonic)
DM EOS, as the horizontal axis of \(\Lambda-M\) relation scales with
\(M_{max}\), which depends on \(\mu\) (\(\rho_0\)). For example, a 2.6
\(M_\odot\) DM star may have a \(\Lambda\) around a few hundreds if the
\(M_{max}\) is around 2.6 \(M_\odot\), but it will become a few thousands if
the \(M_{max}\) is 2.9 \(M_\odot\) instead.

\section{\label{sec_result}Result}
\subsection{\label{sec_general} DM Admixed Neutron Stars with Various DM Mass
Fractions}
\begin{figure}[htb]
\includegraphics[width=\linewidth]{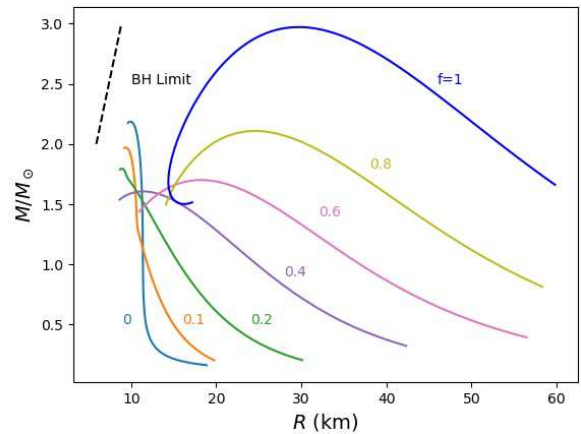}
\caption{\label{fig_APR-0.5-mr} {Mass-radius relations of DM-admixed
compact stars constructed with the APR EOS and \(\mu=\) 0.5 GeV fermionic DM
EOS for different DM fractions \(f\). The dashed line is the black hole limit.
}}
\end{figure}

\begin{figure}[htb]
\includegraphics[width=\linewidth]{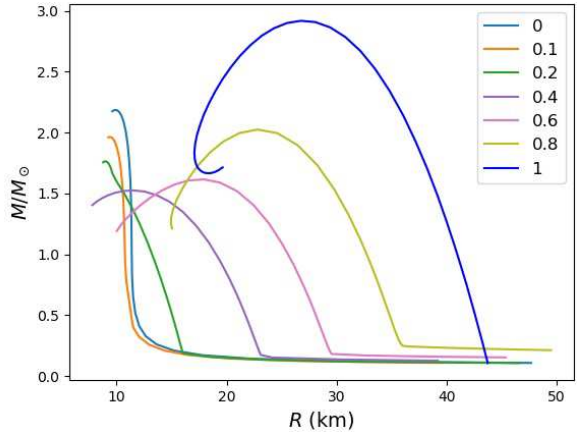}
\caption{\label{fig_APR-1-185-mr} {Same as Fig. \ref{fig_APR-0.5-mr},
but with the APR EOS and \(\rho_0\hbar^3=2.93\times10^{-4}\) GeV\(^4\) bosonic
DM EOS.}}
\end{figure}

After considering our models of pure NM neutron stars, fermionic, and bosonic
DM stars, we now study more generally the properties of DM-admixed compact
stars using a two-fluid description. In Fig. \ref{fig_APR-0.5-mr}, we show the
mass-radius relation of two-fluid stars with different DM mass fractions,
constructed with the APR EOS and 0.5 GeV fermionic DM particle mass. The DM
mass fraction \(f\) is defined as the ratio of the DM mass to the total mass of
the star. The shape of the line for \(f=0.1\) in Fig. \ref{fig_APR-0.5-mr} is
similar to that of a pure NM star, except for a segment showing a different
trend for mass smaller than 1.3 \(M_\odot\). The \(f=0.1\) curve starts to
deviate to a larger radius. This tail behaves more similar to the pure DM
(\(f=1\)) curve, with a more gentle slope. {It is found that kinks on a
curve appear when the two components have the same radius.} This property may
play a role in the tidal properties of a star as they are related to the
compactness of the star \cite{pp_iloveq,pp_tidal_NSGW}, which is the ratio of
the total mass to the outer radius. Similar results are observed when the
bosonic EOS is used (Fig. \ref{fig_APR-1-185-mr}). {In Fig.
\ref{fig_APR-1-185-mr}, there are two kinks for \(f=0.1\) and $0.2$. For 
$f = 0.1$, the two kinks are near \(R=10\) km. For $f=0.2$, one of the kinks is
near $R=10$ km and the other is near $R = 16$ km. The segment in between the
two kinks concaves downward, similar to the pure DM (\(f=1\)) curve, but not
the pure NM (\(f=0\)) curve that concaves upward. The segments separated by a
kink have similar shapes as those of either the pure NM or pure DM limit. A
segment of the mass-radius curve is similar to that of the component with the
larger radius. For larger $f$, only one kink is observed near $M = 0.3
M_{\odot}$ of each curve. The configurations on the flat tails have NM
components that are more extended than the DM. The shapes of the tails are all
similar to that of the pure NM limit, whereas the pure DM case has no flat
tail. This indicates that these flat tails exist because of the extended NM
component. The configurations on the flat tails have very low mass and large
radius, or very low compactness. Therefore, these configurations are not in the
range of our interest even if they are stable. Similar results can also be
observed for other EOSs. We will see later that the relative sizes of the two
components play an important role in admixed stars.}

\begin{figure}[htb]
\includegraphics[width=\linewidth]{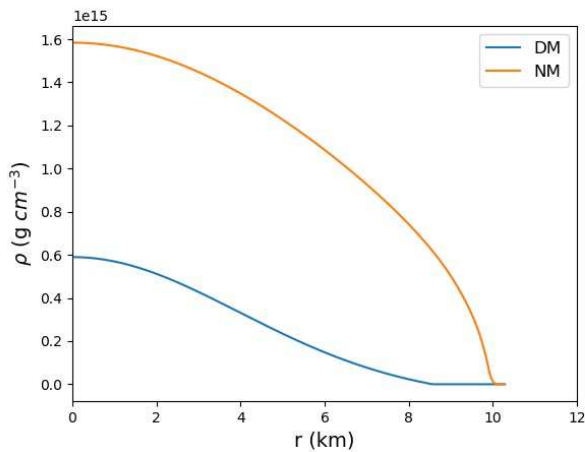}
\caption{\label{fig_APR_1_185_0255_01_density} {Density profile of a
configuration in Fig. \ref{fig_APR-1-185-mr}, where $\beta = 0.255$ and 
$f = 0.1$.}}
\end{figure}

\begin{figure}[htb]
\includegraphics[width=\linewidth]{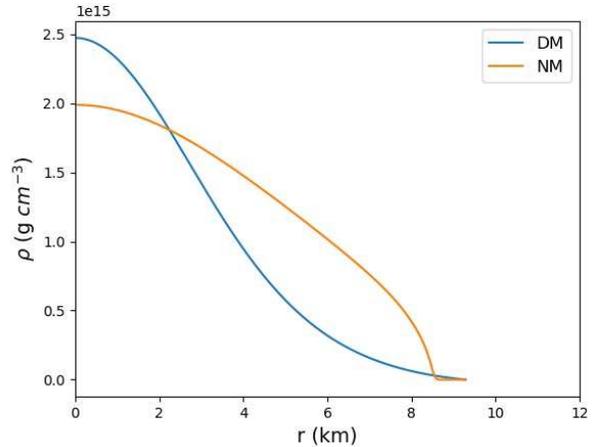}
\caption{\label{fig_APR_1_185_0255_03_density} {Same as Fig.
\ref{fig_APR_1_185_0255_01_density}, but with $f = 0.3$.}}
\end{figure}

{In Figs. \ref{fig_APR_1_185_0255_01_density} and
\ref{fig_APR_1_185_0255_03_density}, we plot the NM and DM density profiles of
two particular star models in Fig. \ref{fig_APR-1-185-mr} as an illustration.
They have the same compactness $\beta = 0.255$ but different $f$. For the model
in Fig. \ref{fig_APR_1_185_0255_01_density}, the DM radius is smaller than the
NM radius and the DM mass only contributes $10\%$ of the total mass. However,
the DM of the model in Fig. \ref{fig_APR_1_185_0255_03_density} is the larger
component and has a higher density near the core. We expect these
configurations to show different tidal properties. The tidal properties of a
star may indicate the existence of a second admixed fluid. }

\begin{figure}[htb]
\includegraphics[width=\linewidth]{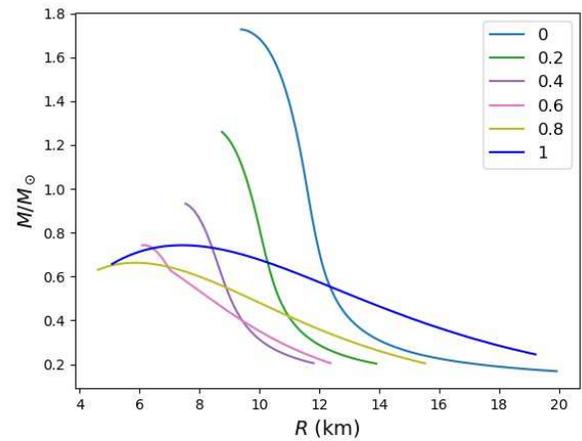}
\caption{\label{fig_LNS-1.0-mr} {Same as Fig. \ref{fig_APR-0.5-mr}, but
with the LNS EOS and \(\mu = \) 1.0 GeV.}}
\end{figure}

\begin{figure}[htb]
\includegraphics[width=\linewidth]{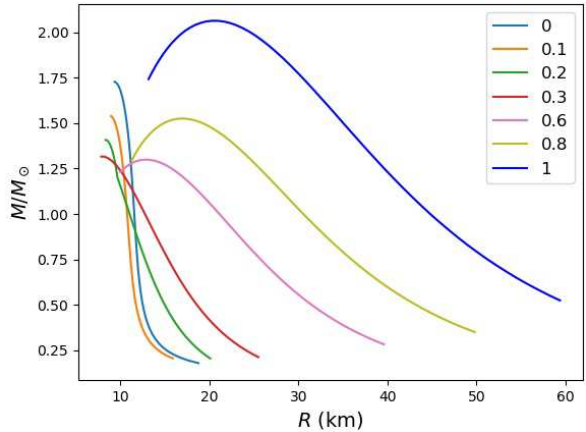}
\caption{\label{fig_LNS-0.6-mr} {Same as Fig. \ref{fig_APR-0.5-mr}, but
with the LNS EOS and \(\mu = \) 0.6 GeV.}}
\end{figure}

In Figs.~\ref{fig_LNS-1.0-mr} and \ref{fig_LNS-0.6-mr}, the mass-radius
relations are generated by the same NM EOS but with different DM EOSs. The NM
EOS is LNS, while the DM EOS is ideal Fermi gas, with DM particle mass of 1.0
GeV (Fig. \ref{fig_LNS-1.0-mr}) or 0.6 GeV (Fig. \ref{fig_LNS-0.6-mr}). The
\(M_{max}\) of pure DM stars for \(\mu=\) 0.6 GeV (1.0 GeV) is greater
(smaller) than that of the NM EOS (see Fig.~\ref{fig_m-r_all}). 
{In both cases, the additional component does not increase \(M_{max}\),
which occurs either in the pure DM or pure NM stars.} This is also true for
other EOSs we have used. Gravity is contributed by both fluids, but the
pressure of each fluid can only support the corresponding fluid itself. It is
not surprising that a two-fluid star cannot support as much total mass as the
one-fluid limit. 

It was suggested in \cite{pp_normal_mode} that the stability of DM-admixed
neutron stars can be deduced from the $M-\rho_c$ relation for a fixed \(f\) in
the same way as for one-fluid stars. The turning point on a given mass-radius
relation represents the maximum stable mass configuration. The stars beyond the
turning point (on the branch of smaller $R$) are unstable against radial
perturbations.  


\begin{figure}[htb]
\includegraphics[width=\linewidth]{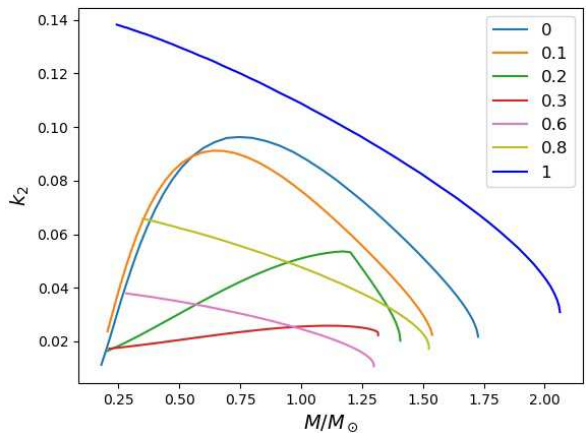}
\caption{\label{fig_LNS-0.6-k2-m} {Same as Fig. \ref{fig_LNS-0.6-mr},
but for the tidal Love number against total mass.}}
\end{figure}

A kink similar to those in Figs. \ref{fig_APR-0.5-mr} and
\ref{fig_APR-1-185-mr} is observed in the tidal Love number against total mass
curve, and is more significant. Fig. \ref{fig_LNS-0.6-k2-m} shows the results
for the LNS EOS with $\mu =0.6$ GeV as an example. We can see that the tidal
Love number may drop to a half or even less as $f$ is increased for a fixed
total mass. {Relations between \(k_2\) and $M$ for neutron stars modeled
by different nuclear matter EOSs were studied in \cite{pp_tidal_NSGW}.} For
$f=0.1$, the change in \(k_2\) relative to a pure NM neutron star is not
significant compared with the differences arising from different neutron star
EOSs. {So, for such a small amount of DM admixed, it would be difficult
to distinguish a DM admixed neutron star from a traditional neutron star
without DM through the tidal Love number.} However, the situation is different
for $f=0.2$. For $M <$ 1.25 \(M_\odot\), $k_2$ decreases significantly compared
to the pure NM result, by more than 50\%. The kink on the \(k_2-M\) relation
induces a large change in \(k_2\), which may be a possible signature of
DM-admixed neutron stars. These kinks will be significant only for some range
of $f$, possibly due to the fact that the two components may have the same
radius only for some $f$. {Similar results can be observed with other
choices of EOS, but the positions of the kinks, the range of $f$ that the kinks
are present and the change in the value of \(k_2\) are sensitive to the EOS. }

\begin{figure}[htb]
\includegraphics[width=\linewidth]{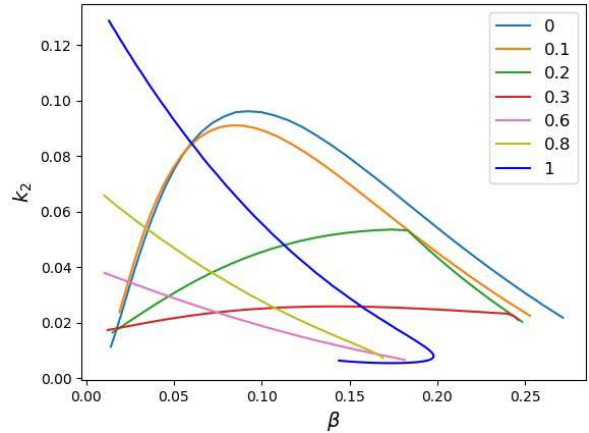}
\caption{\label{fig_LNS-0.6-k2-beta} {Same as Fig.
\ref{fig_LNS-0.6-k2-m}, but for the tidal Love number against compactness.}}
\end{figure}

{It is interesting to understand why and how the tidal Love number
changes when DM is admixed. For better comparison, we plot the tidal Love
number against compactness in Fig. \ref{fig_LNS-0.6-k2-beta}. There are kinks
on the lines with $f = 0.2$ (around $\beta = 0.19$) and $f = 0.3$ (around
$\beta = 0.24$). These kinks are located near the configuration with the same
NM and DM radii. The $k_2-\beta$ curves are similar for $f = 0, 0.1$, and the
right half of $f = 0.2$. Before the DM component takes up a larger radius than
the NM's, the effect of the DM admixture simply shifts the $k_2-\beta$ curve
but preserves its general shape. For $f > 0.5$, the tidal Love number decreases
when the compactness increases monotonically. 
This trend has also been observed for polytropic star models \cite{pp_quark}.
As our DM EOSs are similar to the polytropic EOS, it is not surprising that our
results for high DM fraction show a similar trend.}

\begin{figure}[htb]
\includegraphics[width=\linewidth]{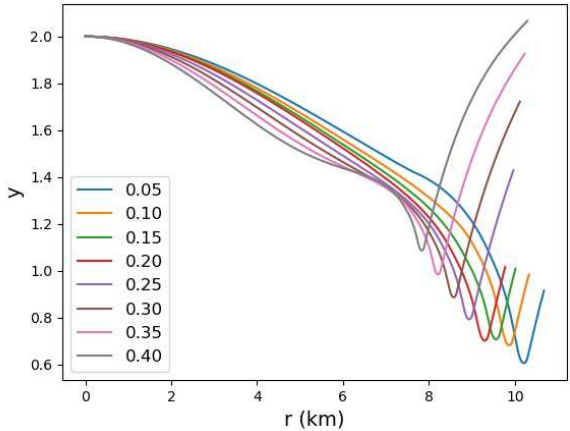}
\caption{\label{fig_LNS-0.6-018-y2_all} {Profiles of $y(r)$ for $\beta =
0.18$ and different $f$ from 0.05 to 0.40. Same setting as
Fig.\ref{fig_LNS-0.6-k2-beta}}}
\end{figure}

\begin{figure}[htb]
\includegraphics[width=\linewidth]{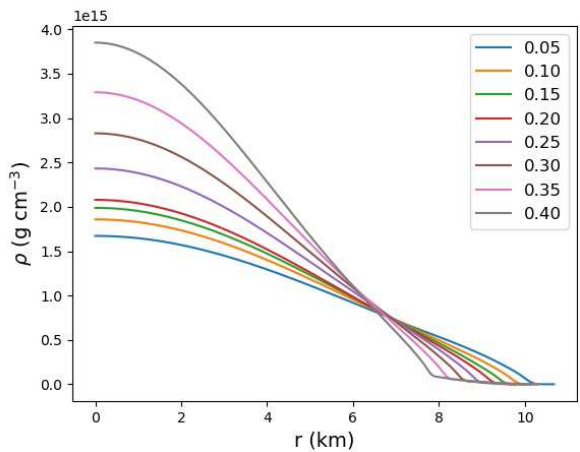}
\caption{\label{fig_LNS-0.6-018-rho_all} {Total energy density profiles
for $\beta = 0.18$. Same setting as Fig.\ref{fig_LNS-0.6-k2-beta}}}
\end{figure}

{Interestingly, when $\beta$ is around 0.05 to 0.20 and $f$ is around
0.2 to 0.4, the tidal Love number is significantly lower than that of the pure
NM case. Several NM EOSs were studied in \cite{pp_quark}, and it was shown that
the tidal Love number for a pure NM neutron star typically peaks at around 0.1
to 0.15. By considering the profile of $y(r)$ defined in Eq. \ref{eq:Love_y}
and its value at the surface $y_R$, we find that the low density region of a
star plays a role in the suppression of the tidal Love number when DM is
admixed. We have studied the configurations around the kink ($\beta = 0.18$) on
the $f = 0.2$ line in Fig. \ref{fig_LNS-0.6-k2-beta}. In Fig.
\ref{fig_LNS-0.6-018-y2_all}, we plot the profiles of $y(r)$ for $\beta = 0.18$
and different $f$. For $f < 0.2$, the v-shape curves are shifted to smaller
radius when $f$ increases, while $y_R$ remain more or less the same. However,
for $f > 0.2$, the v-shape curves have a much longer extension of positive
slope side, and they also shift upwards as $f$ increases. Thus, the values of
$y_R$ for $f>0.2$ are much larger than those of $f < 0.2$. Although the tidal
Love number $k_2(\beta, y_R)$ is a complicated function of $\beta$ and $y_R$,
in the region we are interested in, $k_2$ decreases when $y_R$ increases. Thus,
we get a much lower tidal Love number for $f > 0.2$. Moreover, the v-shape in
the $y(r)$ curve is confined to the low density region of the star. In Fig.
\ref{fig_LNS-0.6-018-rho_all}, we plot the total energy density profiles for
the star models corresponding to the results presented in Fig.
\ref{fig_LNS-0.6-018-y2_all}. It is noted that the minima of the v-shape curves
in Fig. \ref{fig_LNS-0.6-018-y2_all} are located near the positions where the
density is very low and its slope has a drastic change. For $f < 0.2$, the NM
is still the larger component, and the DM component only affects the surface
distribution of NM slightly, resulting in only small changes in $y(r)$, $y_R$,
and therefore $k_2$. For $f > 0.2$, the DM becomes the larger component and
contributes to much larger $y(r)$,  $y_R$, and therefore lower $k_2$. Similar
results are observed by fixing $f$ but varying $\beta$. }

\begin{figure}[tb]
\includegraphics[width=\linewidth]{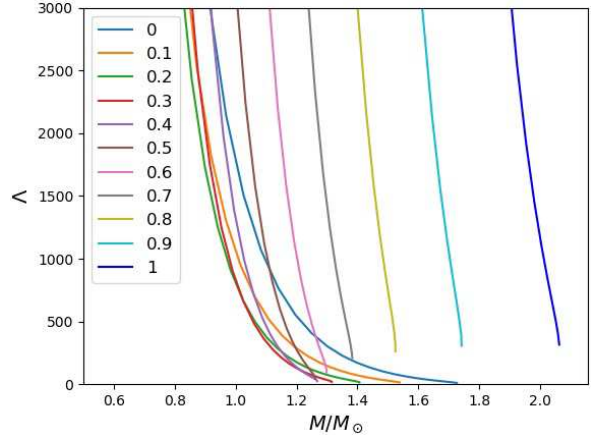}
\caption{\label{fig_LNS-0.6-lambdabar-m} {Same as Fig.
\ref{fig_LNS-0.6-k2-m}, but for the dimensionless tidal deformability.}}
\end{figure}

Let us now focus on the case that the \(M_{max}\) of the DM EOS is larger than
that of the NM EOS. Fig. \ref{fig_LNS-0.6-lambdabar-m} is similar to Fig.
\ref{fig_LNS-0.6-k2-m}, but for the dimensionless tidal deformability againist
total mass. Unlike the previous results in Figs. \ref{fig_APR-0.5-mr} and
\ref{fig_LNS-0.6-k2-m}, the curves are generally smooth. For \(f < 0.2\), the
\(\Lambda-M\) curves are similar to each other.. For example, for \(M=\) 1.25
\(M_\odot\), the dimensionless tidal deformability of the \(f=0.2\) case is
around 70\% smaller than that of a pure neutron star, and is around 85\%
smaller for 1.4 \(M_\odot\). This result agrees with that in
\cite{pp_admixedNS_0.05}, which shows that a \(M=1.4 M_\odot\) neutron star
will have a smaller \(\Lambda\) when a small amount of DM is admixed. The
DM-admixed \(\Lambda-M\) curves are shifted to smaller stellar mass compared
with that of the pure NM case, and thus, \(\Lambda\) is decreased for a fixed
stellar mass but larger \(f\). This seems to be a general property regardless
of the mass of the star.

\begin{figure}[htb]
\includegraphics[width=\linewidth]{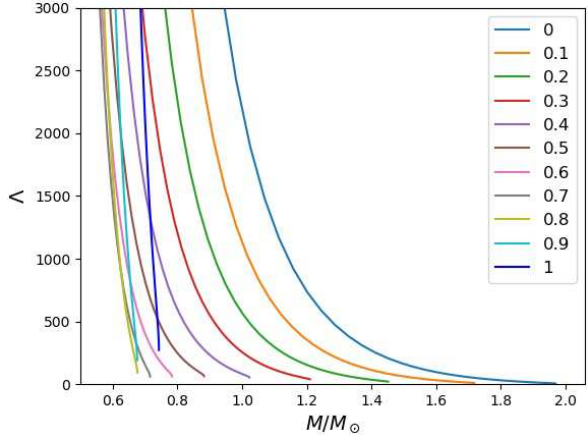}
\caption{\label{fig_KDE0v1-1.0-lambdabar-m} {Same as Fig.
\ref{fig_LNS-0.6-lambdabar-m}, but for the KDE0v1 EOS with \(\mu = 1.0\) GeV.}}
\end{figure}

The dimensionless tidal deformability starts to increase for \(f>0.4\) in Fig.
\ref{fig_LNS-0.6-lambdabar-m}. The curves for \(f=0.5\) to \(0.7\) are steep.
The separations between the curves are larger than those with \(f<0.5\). This
indicates that \(\Lambda\) is very sensitive to $M$ and $f$. \(\Lambda\)
increases rapidly when \(f\) increases in this range. A change in \(f\) will
shift the curve horizontally on the graph, which gives a rapid change in
\(\Lambda\). The large range of possible \(\Lambda\) values may save some NM
EOSs from being ruled out by observations with \(f\) as an extra degree of
freedom. However, it will also be difficult to distinguish and select the NM
EOSs and constrain the DM parameters in this range of DM fractions, for which a
DM halo is formed. A similar rapid increase in \(\Lambda\) is also observed for
the DM halo models studied in \cite{pp_DMhalo}. Qualitatively similar results
can be observed for other choices of the EOS. For example, the APR EOS with 0.4
GeV fermionic DM particle mass shows similar results, but $\Lambda$ starts to
increase at around \(f=0.1\) instead. Note that we have only considered the
cases where the DM EOS has a larger \(M_{max}\) than that of the NM. 
For the opposite situation where the DM EOS has a smaller \(M_{max}\) than that
of the NM EOS, we consider the KDE0v1 EOS with \(\mu=\) 1.0 GeV as an example,
and the corresponding \(\Lambda-M\) relation is shown in Fig.
\ref{fig_KDE0v1-1.0-lambdabar-m}. The curves are almost vertical for 
\(f > 0.5\). The \(\Lambda\) is thus sensitive to $M$. Qualitatively similar
results can be observed for other choices of the EOS.

In all the results we have shown, the properties of two-fluid stars show
continuous change between the limits of pure NM and pure DM stars, with an
abrupt transition at an intermediate DM mass fraction. The \(\Lambda-M\) curves
become steep at some intermediate DM fractions, implying that \(\Lambda\) will
be very sensitive to the \(M\) and \(f\). Also, for DM EOS with smaller
\(M_{max}\), the slope of the \(\Lambda-M\) curve is steeper, so that
\(\Lambda\) is sensitive to \(M\). For DM EOS with larger \(M_{max}\), the
separation between the curves at intermediate DM fractions is large, so that
\(\Lambda\) is sensitive to \(f\).

\begin{figure}[htb]
\includegraphics[width=1.15\linewidth]{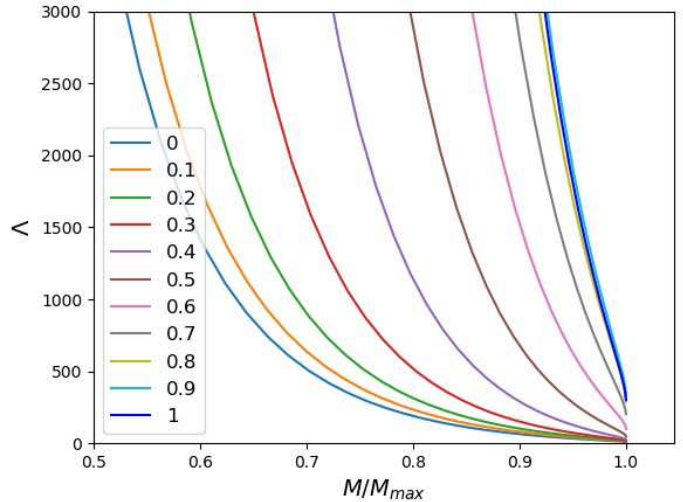}
\caption{\label{fig_LNS-0.6_lambdabar-m_norm}  {Same as Fig.
\ref{fig_LNS-0.6-lambdabar-m}, but with the total mass normalized by
\(M_{max}\) of each curve.}}
\end{figure}

In Fig. \ref{fig_LNS-0.6_lambdabar-m_norm}, we plot $\Lambda$ against
$M/M_{max}$ for different $f$. The normalized relation is less sensitive to 
\(f\), when \(f\) is high ( $> 0.8$ in this example). This result is similar to
the fact that properties of DM stars for the DM EOSs we considered are
self-similar and scale with \(M_{max}\), for different DM parameters. The large
separation between the curves in Fig. \ref{fig_LNS-0.6-lambdabar-m} indicates
that \(\Lambda\) is very sensitive to both \(f\) and \(M\) in this range.
However, we may utilize the result that the \(\Lambda-M\) relations are
self-similar for large \(f\), so that we can reduce the relations to a single
one for $M/M_{max}$. So, we may study the relation between $M_{max}$ and $f$,
instead of that of \(\Lambda\) and \(f\). Also, although $f=0.7$ is not
perfectly fitted, it is still approximately the same as the others, except a
few percentage shift along $M/M_{max}$. Similar behaviour can be observed with
other choice of EOS when the DM EOS has a greater maximum mass than that of NM.

Also, except for $f>0.8$, where the $\Lambda-M/M_{max}$ curves are similar, the
$\Lambda-M/M_{max}$ curves of smaller $f$ are always on the left of those for
higher $f$, and there is no crossing between the curves. This is different from
Fig. \ref{fig_LNS-0.6-lambdabar-m}, where the $\Lambda-M$ curves move back and
forth along the horizontal direction and cross with others. The transition from
pure NM to pure DM is clearer after we normalize $M$ by $M_{max}$. This
suggests that $\Lambda$ should be studied as a function of both $M$ and
$M/M_{max}$.

\subsection{Massive DM-Admixed Neutron Stars}
In the future, more gravitational-wave events similar to GW190814 may be
observed. Although the tidal properties were not measured for the GW190814 
$2.6 M_\odot$ compact object, we will use it as an example to study compact
objects in the mass gap.

The nature of the 2.6 \(M_\odot\) object is still unknown. It may be the
lowest-mass black hole ever observed, or the largest-mass neutron star. The
pure NM neutron stars constructed from the EOSs we use, as well as those from
many other EOSs, cannot reach 2.6 \(M_\odot\). The 2.6 \(M_\odot\) object could
be a DM-admixed neutron star or even a pure DM star, and if so, we may
constrain the range of DM parameters. It is found that even admixed with DM, a
two-fluid star will only reach its maximum mass at either the pure NM or pure
DM limits. So, the DM-admixed neutron star allows a maximum mass of 2.6
\(M_\odot\) only if the DM EOS can reach 2.6 \(M_\odot\). Indeed, \(M_{max}
=2.6 M_\odot\) can be reached if \(\mu <0.535\) GeV for fermionic DM and
\(\rho_0\hbar^3<3.69\times10^{-4}\) GeV\(^4\) for bosonic DM. A much higher
mass limit for the DM EOS can be achieved if we consider a smaller DM particle
mass. However, the radius and \(\Lambda\) of such a DM star will also increase
significantly. {Other constraints may be applied, such as the radius of 
the DM component should be within the binary system, and the star should be
stable against tidal disruption during the inspiral phase.}

Furthermore, if the tidal properties of the binary system are measured, we may
narrow down the DM parameter space. When the DM fraction is high, we have shown
that the \(\Lambda-M\) relations are similar to that of the pure DM stars if
they are normalized by \(M_{max}\). This indicates that they share
approximately the same dimensionless function, i.e. the relations can be
written as
\begin{align}
\Lambda(M/M_{max}(f);\sigma) \approx \Lambda(M/M_{max}(f=1);\sigma),
\end{align}
where $\sigma$ denotes the parameter for the DM EOS. Also, as mentioned, the DM
EOSs we use are self-similar, which means that they share the same
dimensionless function that is independent of the parameter:
\begin{align}
\Lambda(M/M_{max}(f=1);\sigma) = \Lambda(M/M_{max}(f=1)).
\end{align}
Therefore, all these \(\Lambda-M\) relations share approximately the same
dimensionless function. All the information are described by the normalizing 
factor, which is the maximum mass of a \(\Lambda-M\) relation, with a given NM
EOS, fixed DM parameters and a fixed DM fraction $f$. By considering the
maximum mass with different combinations of parameters, we may give constraints
to the parameter space. 

\begin{figure}[htb]
\includegraphics[width=\linewidth]{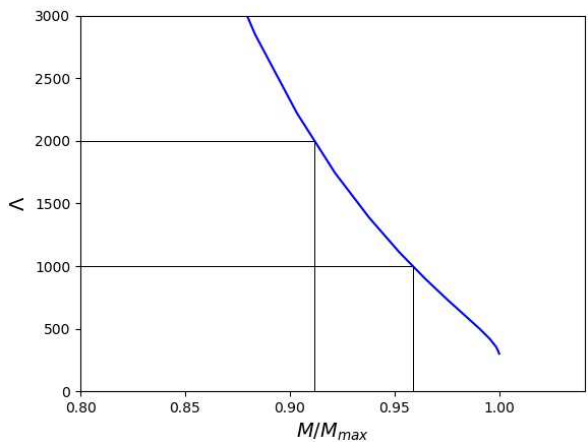}
\caption{\label{fig_lambda-Mbar-boson} $\Lambda - M$ relation for bosonic DM 
EOSs, with $M$ normalized by the maximum mass. Black lines indicate the range 
of variables as the example.}
\end{figure}

\begin{figure}[htb]
\includegraphics[width=\linewidth]{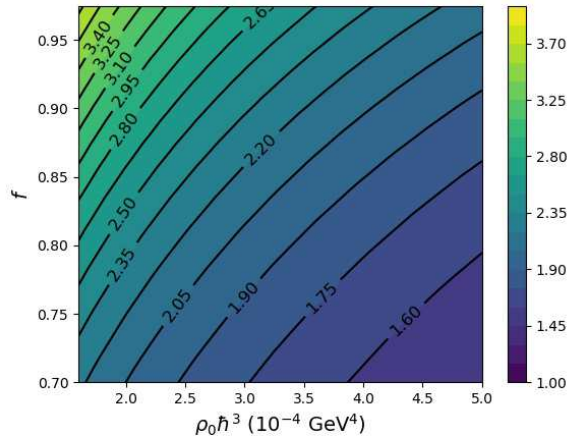}
\caption{\label{fig_M-contour} A contour plot of maximum mass as a function of 
DM fraction and $\rho_0\hbar^3$. APR EOS are assumed for NM, and bosonic DM are
assumed.}
\end{figure}

We demonstrate the approach with an example. Assume we have observed a star 
with mass in between [2.55,2.65] $M_\odot$ with $\Lambda$ in between [1000, 
2000]. We assume NM EOS to be APR EOS and DM EOS bosonic. In Fig. 
\ref{fig_lambda-Mbar-boson}, we can limit the range of the $M/M_{max}$ by 
$\Lambda$. Although the \(\Lambda-M\) relations with high DM fraction is not 
perfectly fitted on the dimensionless form, they behave like shifting along the
axis with a few percentage deviation. So, we may include this approximation in 
the range of mass. Thus, the range of $M/M_{max}$ lies in approximately [0.90, 
0.97], and the maximum mass will be in [2.63, 2.94] $M_\odot$. Fig. 
\ref{fig_M-contour} shows a contour plot for the maximum mass as a function of 
the DM fraction and $\rho_0 \hbar^3$. The parameter space is then constrained. 
Although the NM EOS is still unknown, this approach can be carried out with 
different NM EOSs and then the results combined. This graph only shows a range 
of parameters. It is possible to extend the axis of $\rho_0 \hbar^3$ to even 
lower values, but there may be some constraints as mentioned before. We have 
not ruled out the low DM fraction part, but that is the case that this approach
cannot be directly applied to. {Also, the way to define ``high'' DM 
fraction needs further work.}

\section{\label{sec_discussion}Discussion}

{We have studied the static configurations and tidal properties of DM 
admixed neutron stars. We observe drastic changes (kinks) in the tidal Love 
number as a function of compactness or stellar mass when the NM and DM 
components have the same radius. For small (large) $f$, the tidal Love number 
behaves similar to that of a pure NM (DM) star as expected. However, for 
intermediate values of $f$, such as around 0.3, the tidal Love number is much 
reduced relative to that of a pure NM star. We find that in such cases, the DM
component has a low density tail engulfing the NM component, which leads to a 
significant decrease of the tidal Love number. Also, we have studied the 
dimensionless tidal deformability $\Lambda$. For small $f$, where the star 
configuration is similar to a pure NM star, \(\Lambda\) will tend to decrease 
when more DM are admixed. For large $f$, where the star configuration is 
similar to a pure DM star, the $\Lambda-M$ curves can be scaled to that of the
pure DM stars. Further study about the similarity of $\Lambda (M/M_{Max})$ for
different $f$ may help to relate the properties of pure DM stars to those for 
stars with large $f$. The tidal properties of stars with intermediate DM 
fractions are much more sensitive to the DM parameters. }

{The existence of the DM component hardly helps to increase the total 
mass of the star unless the DM fraction is high. However, this means that the 
two-fluid star is more like a DM star instead of a neutron star. A pure DM 
star can have a \(M_{max}\) larger than that of the two-fluid stars. 
Therefore, we may make use of massive compact object in GW190814 \cite{pp_26} 
as DM-admixed stars to limit the DM parameter space, if such a star is 
believed to have a high DM fraction. If the recently discovered 2.6 
\(M_\odot\) compact object is a DM-admixed neutron star with a high DM 
fraction, the fermionic DM would have \(\mu < 0.535\) GeV, and the 
self-interacting bosonic DM would have \(\rho_0\hbar^3 < 3.69\times10^{-4}\) 
GeV\(^4\). Any more massive compact objects that are not black holes, if 
detected, will give even tighter constraints on these DM parameters, provided 
that these star have a high DM fraction. For compact objects in the mass gap, 
we have also illustrated a method to limit the DM parameters and DM fraction 
if the DM fraction is high.}

\section*{Acknowledgments}
This work is partially supported by a grant from the Research Grants Council of 
the Hong Kong Special Administrative Region (Project No. 14300320).

\appendix
\section{\label{sec_TOV}{Derivation from General Relativistic Two-fluid 
Formalism}}
{We follow the general relativistic two-fluid formalism used in 
\cite{Comer1999} and \cite{pp_normal_mode}. We will use a similar notation as 
\cite{pp_normal_mode}, except that the number density current for DM will be 
denoted as $d^\alpha$ and the master function will be denoted as $\Phi$. The 
master function plays the role of EOS in the two-fluid formalism and is defined
by the number density currents of the two fluids as discussed below. The 
Einstein field equation and the hydrodynamics equations reduce to the following
equations by considering a static and spherically symmetric spacetime} 
\cite{pp_normal_mode} ,
\begin{align}
\lambda' = \frac{1-e^{\lambda}}{r} - 8 \pi r e^{\lambda} \Phi,
\end{align}
\begin{align}
\nu'= -\frac{1-e^{\lambda}}{r} + 8 \pi r e^{\lambda}\Psi,
\end{align}
\begin{align}
A^0_0 d' + B^0_0 n' + \frac{1}{2}(Bn+Ad)\nu'=0, \label{eq:A3}
\end{align}
\begin{align}
C^0_0 d' + A^0_0 n' + \frac{1}{2}(An+Cd)\nu'=0, \label{eq:A4}
\end{align}
{where the prime denotes the derivative with respect to $r$, and}
\begin{align}
A = - \frac{\partial \Phi}{\partial (x^2)}, B = - 2 \frac{\partial
\Phi}{\partial (n^2)}, C = - 2\frac{\partial \Phi}{\partial (d^2)},
\end{align}
\begin{align}
A^0_0 = A + 2 \frac{\partial B}{\partial (d^2)} nd + 2 \frac{\partial
A}{\partial (n^2)} n^2 + 2 \frac{\partial A}{\partial (d^2)} d^2 +
\frac{\partial A}{\partial (x^2)} dn,
\end{align}
\begin{align}
B^0_0 = B + 2 \frac{\partial B}{\partial (n^2)} n^2 + 4 \frac{\partial
A}{\partial (n^2)} nd + \frac{\partial A}{\partial (x^2)} d^2,
\end{align}
\begin{align}
C^0_0 = C + 2 \frac{\partial C}{\partial (d^2)} d^2 + 4 \frac{\partial
A}{\partial (d^2)} nd + \frac{\partial A}{\partial (x^2)} n^2,
\end{align}
{where $n^2$, $d^2$, and $x^2$ are scalars defined by the NM $n^\mu$ and
DM $d^\mu$ number density currents:}
\begin{align}
n^2 = -n_\alpha n^\alpha, d^2 = -d_\alpha d^\alpha, x^2 = -n_\alpha d^\alpha.
\end{align}
{The master function $\Phi$ is in general a function of $n^2$, $d^2$,
and $x^2$. The generalized pressure $\Psi$ is given by }
\begin{align}
\Psi = \Phi + \mu n + \chi d,
\end{align}
{where $\mu = B n + A d$ and $\chi = Cd + A n$ are the chemical
potentials of NM and DM, respectively. With a given master function and
suitable boundary conditions, the above equations can be used to construct a
non-rotating two-fluid star in general relativity \cite{Comer1999,
pp_normal_mode}.}

{Now, we make the assumption that NM and DM only interact with each
other through gravity. This means that the two fluids affect each other only
through the effect of the metric. This assumption means that the master
function $\Phi$ does not depend on the cross term $x^2$ so that $\Phi$ can be
separated into two parts, }
\begin{align}
\Phi(n^2,d^2,x^2) = \Phi_n(n^2) + \Phi_d(d^2).
\end{align}
{With this assumption, many of the above coefficients can be
simplified:}
\begin{align}
A = A^0_0 = 0, \label{eq:simplified_begein}
\end{align}
\begin{align}
B = - \frac{1}{n}\frac{\partial \Phi_n}{\partial n}, C = -
\frac{1}{d}\frac{\partial \Phi_d}{\partial d}, \label{eq:BC_simplifed}
\end{align}
\begin{align}
B^0_0 = B + \frac{\partial B}{\partial n}n, C^0_0 = C + \frac{\partial
C}{\partial d}d.
\end{align}
{The generalized pressure $\Psi$ can also be separated into two parts as
$\Psi = \Psi_n(n^2) + \Psi_d(d^2)$, where}
\begin{align}
\Psi_n = \Phi_n(n^2) + Bn^2, \label{eq:Psi-Phi}
\end{align}
\begin{align}
\Psi_d = \Phi_d(d^2) + Cn^2.
\end{align}
{It is noticed that}
\begin{align}
\frac{\partial \Psi_n}{\partial n} &= \frac{\partial (\Phi_n + Bn^2) }{\partial
n} = Bn + \frac{\partial B}{\partial n} n^2 = B^0_0n. \label{eq:simplified_end}
\end{align}
{We can substitute Eq.s \ref{eq:simplified_begein} to
\ref{eq:simplified_end} into Eq. \ref{eq:A3} to get}
\begin{align}
\frac{d\Psi_n}{dr} = -\frac{1}{2}(-\Phi_n+\Psi_n)\nu'.
\end{align}
{Same result is also obtained for the DM part. By setting one of the 
components to have zero contribution, the standard TOV equation shall be 
obtained. We shall replace the generalized pressure $\Psi_i$ by the usual 
pressure $p_i$, and master function $\Phi_i$ by the minus of energy density 
$-\rho_i$. The set of two-fluid equations is then obtained in the form of the 
standard TOV equation. We can also check the relation between $\Phi_i$ and 
$\Psi_i$ to see if they fulfill the same relation between the energy density 
and pressure. From thermodynamics, we have the following relation,}
\begin{align}
p_n = -\frac{\partial (\rho_n/n)}{\partial(1/n)} = \frac{\partial 
\rho_n}{\partial n} n - \rho_n.
\end{align}
{From Eq.s \ref{eq:BC_simplifed} and \ref{eq:Psi-Phi}, we have}
\begin{align}
\Psi_n = -(-\Phi_n) + \frac{\partial (-\Phi_n)}{\partial n}n. 
\end{align}
{Similar results can be obtained for both fluids.}

{To compute the tidal Love number, we follow the method in 
\cite{pp_tidal_super, Yeung_2021}. The modification for Eq. \ref{eq_Q} is}
\begin{align}
\frac{\rho + p}{dp/d\rho} \to &-\frac{\mu^2C^0_0+\chi^2 B^0_0-2\mu\chi 
A^0_0}{{A^0_0}^2-B^0_0C^0_0} \nonumber\\
= &\frac{\mu^2}{B^0_0}+\frac{\chi^2}{C^0_0} \nonumber\\
= &\sum_i\frac{\rho_i + p_i}{dp_i/d\rho_i}.
\end{align}

\nocite{*}

\bibliography{main}

\end{document}